\documentclass[preprint2]{aastex}
\usepackage{graphics}
\usepackage{epsfig}
\usepackage{longtable}
\def\gsim{\lower 2pt \hbox{$\, \buildrel {\scriptstyle >}\over
{\scriptstyle \sim}\,$}}
\def\lsim{\lower 2pt \hbox{$\, \buildrel {\scriptstyle <}\over
{\scriptstyle \sim}\,$}}

\newcommand{\xmm}{{\em XMM-Newton}}

\def\HI{H{\small I}\ }

\def\cor{\widehat=}

\shortauthors{}
\shorttitle{}

\begin{document}

\title{An \xmm\ Observation of the Massive Edge-on Sb Galaxy NGC~2613}
\author{
 Z. Li\altaffilmark{1},
 Q. D. Wang\altaffilmark{1}, J. A., Irwin\altaffilmark{2},  \& T. Chaves\altaffilmark{2}
}
\altaffiltext{1}{Department of Astronomy, B619E-LGRT,
       University of Massachusetts, Amherst, MA~01003}
\altaffiltext{2}{Department of Physics, Engineering Physics \& Astronomy, Queen's University, Kingston, ON K7L 3N6, Canada}

\begin{abstract}
We present an \xmm\ observation of the massive edge-on Sb galaxy NGC~2613. We
discover that this galaxy contains a deeply embedded active nucleus with 
a 0.3-10 keV luminosity of 3.3$\times10^{40}{\rm~ergs~s^{-1}}$ and a line-of-sight absorption 
column of $1.2 \times 10^{23} {\rm~cm^{-2}}$. Within the $25{\rm~mag~arcsec^{-2}}$ optical B-band 
isophote of the galaxy, we detect an additional 4 sources with an accumulated 
luminosity of 4.3$\times10^{39}{\rm~ergs~s^{-1}}$. 
The bulk of the unresolved X-ray
emission spatially follows the near-infrared (NIR) K-band surface brightness distribution;the luminosity ratio $L_X/L_K \sim 8\times10^{-4}$ is consistent with
that inferred from galactic discrete sources. 
This X-ray-NIR association 
and the compatibility of the X-ray spectral fit with the
expected spectrum of a population of discrete sources suggest that
low-mass X-ray binaries (LMXBs) are the most likely emitters of the unresolved
emission in the disk region.
The remaining unresolved
emission is primarily due to extraplanar hot gas.
The luminosity of this gas is at least a factor of 10
less than that predicted by recent simulations of intergalactic gas accretion 
by such a massive galaxy with a circular rotation speed 
$V_c \sim 304 {\rm~km~s^{-1}}$ (Toft et al.~2002). 
Instead, we find that the extraplanar hot gas most likely represents
discrete extensions away from the disk,
including two ``bubble-like'' features on either side of the nucleus.
These extensions appear to correlate with radio continuum emission
and, energetically, can be easily explained by outflows from the galactic disk.
\end{abstract}

\section{Introduction}
X-ray observations of extraplanar hot gas ($T$ $\gtrsim$ 10$^6$ K) around nearby edge-on disk galaxies 
are essential in the study of the galactic ecosystem in many aspects, particularly
the disk-halo interaction. 
Such observations have helped establish the prevalence of   
galactic superwinds in starburst galaxies, e.g., 
NGC~253 (Strickland et al.~2000, 2002) and NGC 4666 (Dahlem, Weaver \& Heckman~1998), among others.
Extraplanar X-ray-emitting gas has also been detected unambiguously around 
several ``normal'' late-type galaxies with little evidence for nuclear starbursts:~NGC~891 (Sb; Bregman \& Houck 1997),
NGC 4631 (Scd; Wang et al.~2001), NGC 3556 (Sc; Wang, Chaves \& Irwin 2003) and
NGC 4634 (Scd; T$\ddot{\rm u}$llmann et al.~2006).
In these galaxies (except for NGC~4634 which currently lacks direct evidence), 
extraplanar hot gas is clearly
linked to outflows from recent massive star-forming regions in galactic disks.
The global X-ray properties of extraplanar gas 
in these ``normal'' star-forming galaxies,
when scaled with the star formation rate of the host galaxies,
appear similar to those found in starburst galaxies (Strickland et al.~2004a, b; Wang 2005).
Nevertheless, this needs to be confirmed 
by extended X-ray observations of ``normal'' star-forming galaxies.

On the other hand, current galaxy formation models also predict the existence
of hot gaseous halos surrounding present-day disk galaxies, which arise
from gravitational infall from the intergalactic medium (IGM; e.g.,
Toft et al.~2002 and references therein). The predicted extraplanar X-ray luminosity 
strongly depends on the mass of the host galaxy.
X-ray observations thus have long been 
expected to detect such gaseous halos around nearby massive, typically earlier-type disk galaxies.
However, there is so far little 
direct observational evidence for the presence of this kind of X-ray-emitting 
halo. Benson et al.~(2000)
analyzed X-ray emission from the outer halos ($\gtrsim 5'$) of primarily 
two early-type spirals NGC 2841 
(Sb; $\sim 15$ Mpc) and NGC 4594 (Sa; $\sim 25$ Mpc), using {\sl ROSAT} PSPC
observations. No significant diffuse emission was detected, although
the upper limits to the diffuse X-ray luminosities are
consistent with the current predictions (Toft et al.~2002). 
Therefore, more dedicated searches for the X-ray signals of IGM accretion 
around disk galaxies are needed.

Here we present a study of an \xmm ~observation toward NGC~2613, an edge-on Sb galaxy
with ``normal'' star formation.
We focus on probing the spatial and spectral properties of its large-scale X-ray emission.
This galaxy (Table~\ref{tab:N2613}) is a good candidate to 
probe the presence of hot gas, in the sense that: 1) it is very massive and thus expected
to contain a large amount of hot gas; 2) its high inclination ($\sim$79$^\circ$)
allows the possibility of detecting 
extraplanar emission, either from a halo of accreted gas or a large-scale outflow;
3) its moderately large distance (25.9 Mpc) places the galaxy and its $\sim$50 kpc vicinity
in the field of view (FOV) of a typical \xmm ~observation, offering a good opportunity
of studying the large-scale distribution of gas, and
4) it is known to show extraplanar features at other wavebands, specifically
the radio continuum and \HI
(Chaves \& Irwin 2001; Irwin \& Chaves 2003) as well as earlier \HI
observations (Bottema 1989).

\section{Observations and Data Reduction}
We obtained two \xmm ~observations on NGC~2613. The first observation (Obs.~ID 0149160101), 
taken on April 23/24, 2003 with a 40 ks exposure, suffered heavily from background flares. 
Consequently, another observation (Obs.~ID 0149160201) with a 33 ks exposure was taken 
on May 20/21, 2003. In our analysis, we only used data obtained from the second observation.

We used SAS, version 6.1.0, together with the latest calibration files for data reduction. 
For the MOS data, we selected only events with patterns 0 through 12 and applied flag filters
XMMEA\_EM, XMMEA\_2, XMMEA\_3 and XMMEA\_11. For the PN data, we selected only events with patterns 0 through 
4 and applied flag filters XMMEA\_EP, XMMEA\_2, XMMEA\_3 and XMMEA\_11.
According to light curves of the MOS and the PN, 
we further excluded time intervals with high background rates by setting good time interval
thresolds of 2.0 cts/s for the MOS in the 0.3-12 keV range and 6.0 cts/s for the PN in 
the 0.3-15 keV range, respectively. 
The resulting net exposure time is 23.5 ks for the MOS and 18.7 ks for the PN.
We used the {\sl skycast} program to generate the ``blank-sky'' background for our
observation. Same event filters were applied to the``blank-sky'' event files, resulting in exposures
of 791, 759 and 294 ks for the MOS1, MOS2 and PN, respectively.
We then constructed images and exposure maps at the 0.2-0.5, 0.5-1, 1-2, 2-4.5 and 4.5-7.5 keV bands
for each detector.
For spectral analysis, we selected only events with FLAG = 0.


\section{Analysis and Results} {\label{sec:analysis}}
\subsection{Discrete X-ray Sources} {\label{subsec:ps}}
We perform source detection on the PN images of the soft (S, 0.5-2 keV), hard (H, 2-7.5 keV) and 
broad (B=S+H) bands.
As detailed in Wang (2004), the source detection procedure, optimized to detect point-like sources, 
uses a combination
of detection algorithms: wavelet, sliding-box and maximum likelihood centroid fitting.
The source detection uses a detection aperture of the 50\% PSF encircled energy radius (EER).
Multiple detections with overlapping 2$\sigma$ centroid error circles are considered to
be the same source, and the centroid position with the smallest error is adopted.
The accepted sources all have a local false detection probability $P \leq 10^{-7}$.

\begin{figure}[!htb]
 \vskip -1cm 
 \centerline{
      \epsfig{figure=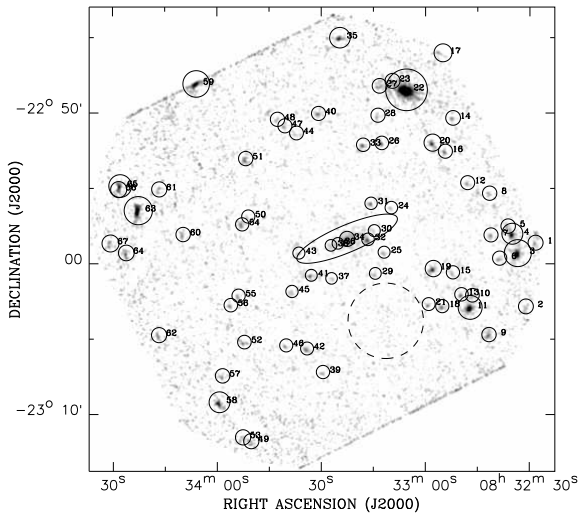,width=0.5\textwidth,angle=0}
    }
  \caption{EPIC-PN intensity image in the 0.5-7.5 keV band after 
flat-fielding. An adaptively
smoothed background has been subtracted from
the image to highlight the discrete sources,
which are marked by circles with radii of twice the 50\% EER. 
The source numbers (Table~{\ref{tab:pn_source_list}}) are also marked. 
The ellipse illustrates the optical $I_B=25{\rm~mag~arcsec^{-2}}$ isophote ($7\farcm2 \times 1\farcm8$) of NGC~2613.
The dashed circle outlines the region where the local background spectrum is extracted for spectral
analysis.}
 \label{fig:sou_plot}
\end{figure}

We detect a total of 67 discrete sources on the PN images,
5 of which are located within the $I_B=25{\rm~mag~arcsec^{-2}}$ isophote of the galaxy.
We have also examined the source detection in the 0.2-0.5 keV band, but found that the results
are contaminated by many artifacts, especially near the CCD boundaries.
Using the same detection procedure, we detected 46 sources in the combined images of
MOS1 and MOS2, but no new source is found within the FOV of the PN.
Table~{\ref{tab:pn_source_list}} lists the sources detected from the PN data,   
the locations of which are shown in Fig.~{\ref{fig:sou_plot}}.

\begin{figure}[!htb]
  \centerline{
    \epsfig{figure=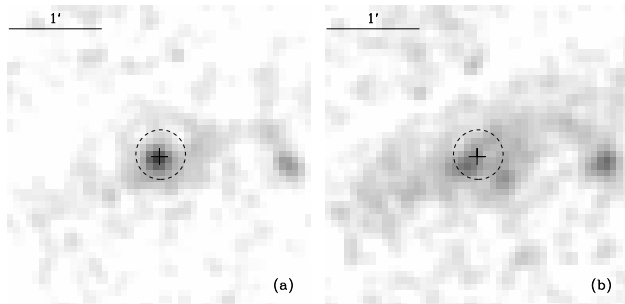,width=0.5\textwidth,angle=0}
    }
  \caption{EPIC-PN intensity images around the galactic center in the 2-7.5 (a) and 0.5-2 (b) keV bands. 
  The nucleus is prominent only in the hard band. 
The plus sign marks the optical center of NGC~2613.
The dashed circle with a 16$^{\prime\prime}$ radius illustrates the region for 
the spectrum extraction.}
 \label{fig:nucleus_plot}
\end{figure}

The prominent nucleus of NGC~2613 is readily seen in Fig.~\ref{fig:nucleus_plot}a.
The nucleus is the only source detected within the central 30$^{\prime\prime}$,
suggesting that this emission may represent an AGN.  
To investigate this further,
we perform a spectral analysis on the nuclear emission.
Due to the relatively low spatial resolution and short exposure time of the observation as well as 
the relatively large distance to NGC~2613, the
spectrum extraction of the nucleus is a compromise between having better counting statistics and suffering 
less contamination from non-nuclear emission around the nucleus.
To assess the contribution from non-nuclear emission,
we extract two spectra from circles with radii of 10$^{\prime\prime}$ 
and 16$^{\prime\prime}$ around the galactic center, for each detector.
The 10$^{\prime\prime}$ (16$^{\prime\prime}$) radius corresponds to a physical scale of 
$\sim$1.2 (2.0) kpc at the distance of NGC~2613 and represents 
an enclosed energy fraction (EEF) of $\sim$0.50 (0.65) 
in the PN and $\sim$0.55 (0.70) in the MOS. 
We extract a background spectrum for each detector from a circle with a 2\farcm5
radius at $\sim6^\prime$ to the south of the galactic disk (Fig.~\ref{fig:sou_plot}). 
This background region is chosen because it shows a low X-ray intensity
and is at an off-axis angular distance comparable to that of the disk.
We then bin the source spectra to achieve a background-subtracted 
signal-to-noise ratio greater than 2.

\begin{figure}[!htb]
  \vskip -1cm
  \centerline{
      \epsfig{figure=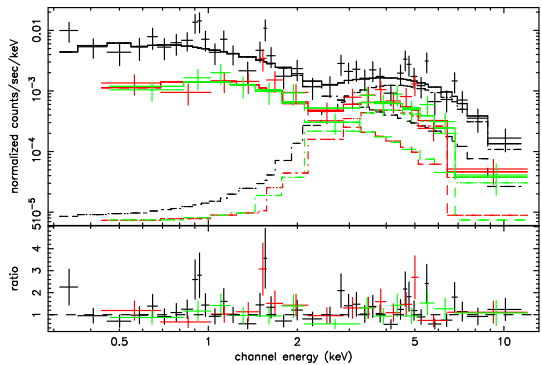,width=0.4\textwidth,angle=270}
    }
  \vskip -0.5cm
  \caption{EPIC spectra (Black: PN spectrum; Red: MOS1 spectrum; Green: MOS2 spectrum)
  of the nucleus of NGC~2613 extracted from a 16$^{\prime\prime}$ circle. 
  The best-fit model (solid curve) consisting of two absorbed power-law components 
  (dashed and dash-dot curves, respectively) is also plotted. The lower panel shows
  the data-to-model ratios.
   }
 \label{fig:nucleus_spec}
\end{figure}

The spectra
from the 16$^{\prime\prime}$ circle show a prominent broad 
bump at energies $\sim$~4-5 keV (Fig.~\ref{fig:nucleus_spec}), which can also be seen in the 
spectra from the 10$^{\prime\prime}$ circle. This is further evidence 
that the nuclear region contains an AGN since the spectra
can naturally be explained by a combination of a heavily absorbed power-law 
from an AGN and a softer contribution from non-nuclear emission.
We jointly fit the PN, MOS1 and MOS2 spectra in the 0.3-12 keV range with XSPEC.
Owing to the lack of obvious features below 3 keV in the spectra, 
a composite model of {\sl wabs[zwabs(PL)+PL]} is applied,
where the first power-law component (PL1)
 with intrinsic absorption characterizes the nuclear emission and 
the second power-law component (PL2)
represents the non-nuclear emission. In the fit,
we require that the amount of foreground absorption be at least that supplied by the
Galactic foreground  
(as specified in {\sl wabs}: N$_{\HI} \geq 6.8{\times}10^{20}{\rm~cm^{-2}}$). 
Fit results for the 10$^{\prime\prime}$ and 16$^{\prime\prime}$ spectra are consistent with each other
within the uncertainty ranges.
We list in Table~\ref{tab:nucleus_fit} the fit results to the 16$^{\prime\prime}$ spectra,
the implications of which will be discussed in \S~\ref{subsec:nucleus}.
All quoted errors in the tables are at the 90\% confidence level.
The best-fit two-component model is shown in Fig.~\ref{fig:nucleus_spec}.

We note that in the PN spectrum
there is some hint of excess over the best-fit two-component model 
at $\sim$0.9 keV,
which might be physically due to the presence 
of diffuse hot gas around the nucleus. 
We thus try to include a thermal plasma component (APEC in XSPEC)
in the fit to probe the existence of an additional thermal component.
While the best-fit temperature of this component is, as expected, $\sim$0.9 keV, 
the fit is not significantly improved according to an F-test, and the range of the fitted temperature
could not be well constrained. 
Therefore we consider this potential thermal component insignificant in the spectra.

Given the relatively high source detection limit ($\sim 2\times10^{38}{\rm~ergs~s^{-1}}$) and the 
relatively low spatial resolution, the bulk of expected galactic X-ray sources of NGC~2613 are still
embedded in unresolved emission.
Indeed, only four sources (in addition to the nucleus) 
are detected within the $D_{25}$ ellipse of the galaxy.
We jointly fit the accumulated spectra of these four sources extracted from individual detectors.
Circles with radii of twice the 50\% EER around
individual sources are used to accumulate the spectra. 
Corresponding background spectra are extracted from the source-removed $D_{25}$ ellipse.  
We find that an absorbed power-low model fits the spectra well, 
giving a best-fit photon index of $1.80^{+0.30}_{-0.27}$
and an intrinsic 0.3-10 keV luminosity of $4.3\times10^{39}{\rm~ergs~s^{-1}}$.
The slope of the power-law is typical for X-ray binaries.
Given their high luminosities ($\gtrsim~10^{39}{\rm~ergs~s^{-1}}$), on average, 
these sources are likely ultraluminous X-ray binaries.
 
\subsection{Unresolved X-ray emission} {\label{subsec:diffuse}}
\subsubsection{Spatial distribution} 
\begin{figure}[!htb]
  \vskip -1cm
   \centerline{
      \epsfig{figure=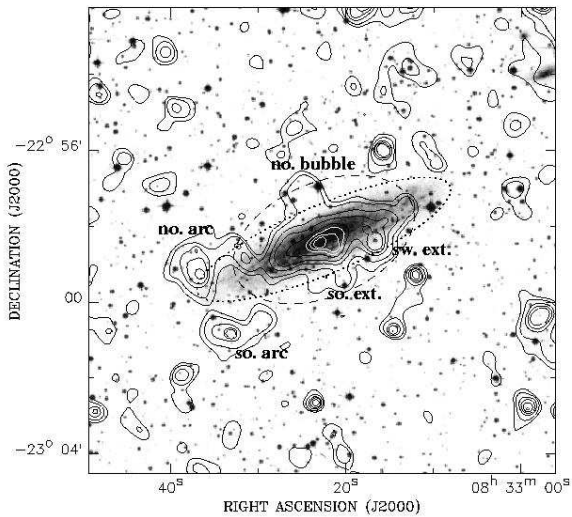,width=0.5\textwidth,angle=0}
    }
  \caption{EPIC-PN 0.5-2 keV intensity contours overlaid on the 
digitized sky-survey (first generation) image of NGC~2613. 
The X-ray intensity is adaptively smoothed with CIAO {\sl csmooth} 
to achieve a signal-to-noise ratio of $\sim$3. 
The contour levels are at 4.1, 5.0, 7.0, 8.8, 13, 16, 22 and 32 $\times$10$^{-3}{\rm~cts~s^{-1}~arcmin^{-2}}$.
The dotted ellipse represents the optical $I_B=25{\rm~mag~arcsec^{-2}}$ isophote of the galaxy.
The dashed ellipse (5$^\prime$ $\times$ 3$^\prime$) illustrates the region where spectra of unresolved X-ray emission are extracted. 
}
 \label{fig:2dmap}
\end{figure}

{\label{subsubsec:spat_anal}}
As an overview, Fig.~\ref{fig:2dmap} shows the large scale 0.5-2 keV X-ray emission 
around NGC~2613 and its similarity with the optical disk.
Along the major axis of the disk, the emission appears rather smooth and 
is confined within $\sim$2\farcm5 from the
galactic center. Whereas along the direction perpendicular to the disk,
some extended features are present, forming well-defined structures.
To the north of the disk 
(Fig.~\ref{fig:2dmap}) is a ``bubble-like'' feature, which is referred to as the
`north bubble' in the following.  
This feature follows the minor axis fairly well and 
has a maximum extent of $\sim100^{\prime\prime}$ (13 kpc) from the nucleus.
Immediately on the opposite
side of the nucleus is another extension that is somewhat smaller, reaching
 $\sim1^\prime$.  
This feature will be called the `south extension'. A third feature
protrudes from the south side of the major axis but west of the minor axis.  This feature
will be called the `south-west feature'.  
Finally two very large extensions are seen to the east of the major
axis with emission peaks located at 
RA= $08^h33^m 37^s$,  DEC$= -22^\circ 59^\prime 19^{\prime\prime}$ (north arc), and
RA= $08^h 33^m 33^s$,  DEC$= -23^\circ 0^\prime 51^{\prime\prime}$ (south arc).
These 
have the appearance of arising from the eastern tip of the X-ray disk and will
be called the `eastern extensions' consisting of northern and southern arcs.
The above features
are labelled in Fig.~\ref{fig:2dmap} for ease of reference.
Most of these features extend significantly beyond the optical disk.
Given the extent of the X-ray emission along the major axis and the inclination of the disk,
the projected in-disk emission along the minor axis should be within $45^{\prime\prime}$,
whereas most of the extended features show an extent larger than 1$^\prime$.
Therefore, we suggest that these features are truly extraplanar.
Their origin will be further discussed in \S~\ref{sec:discussion}.

We remove the detected discrete sources, except for the nucleus,
to study the spatial properties of the unresolved soft X-ray emission.
The nucleus is kept because it is heavily obscured
(\S~\ref{subsec:ps}) and contributes few photons to the emission below 2 keV.
For each faint source with a count rate ($CR$) $\leq$ 0.01 ${\rm cts~s^{-1}}$, 
we exclude a circular region with a radius
of twice the 50\% enclosed energy radius (EER). 
For sources with $CR$ $>$ 0.01 ${\rm cts~s^{-1}}$, 
the radius is multiplied by
an additional factor of 1+log($CR$/0.01).
The choice of the adopted radii is a compromise between removing the bulk of the source contribution
and preserving a sufficient field for the study of unresolved emission. 
Thus 75-80\% of photons from each detected source is removed according to our criteria.

\begin{figure}[!htb]
  \centerline{
      \epsfig{figure=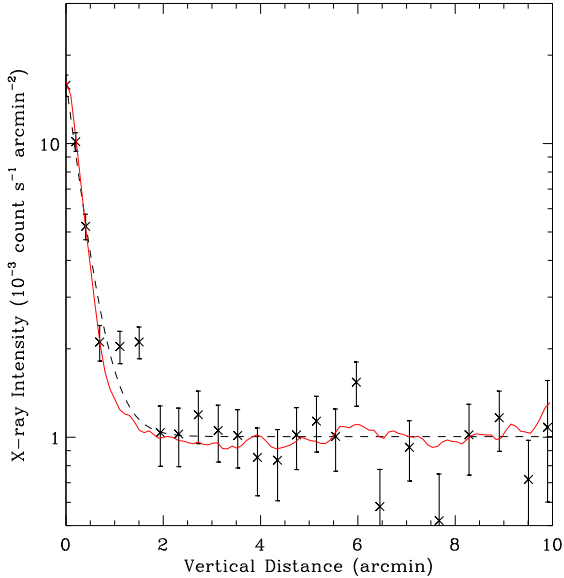,width=0.5\textwidth,angle=0}
    }
  \caption{EPIC-PN 0.5-2 keV intensity profile (black crosses) along the  
direction perpendicular to the disk of NGC~2613.
A ``blank-sky'' background has been subtracted and discrete
sources have been removed, except for the nucleus.
The full width used for averaging the intensity is 5$^{\prime}$.  
Spatial binning is adaptively adjusted to achieve a count-to-noise ratio greater than
12, with a minimum step size of 12$^{\prime\prime}$. 
The black dashed curve is a fit to the profile using 
an exponential law plus a constant local background. 
The red solid curve shows the 2MASS K-band profile convolved with
the PN PSF and 
normalized by a factor of 3.0$\times10^{-3}{\rm~cts~s^{-1}~arcmin^{-2}}/({\rm MJy~sr^{-1}})$ (see text).
A constant of $10^{-3}{\rm~cts~s^{-1}~arcmin^{-2}}$ has been added
to all data points to avoid negative values improper for a logarithmic plot.} 
\label{fig:i_vertical}
\end{figure}

We construct the ``blank-sky'' background-subtracted, exposure-corrected 
0.5-2 keV intensity profile from the source-removed PN image,
as a function of off-center distance along the minor axis.  
The full width along the direction parallel to the major axis used for
averaging the intensity is 5$^{\prime}$ ($\sim$ 0.7 $D_{25}$), 
approximately the maximal extent of the unresolved emission along the major axis (Fig.~\ref{fig:2dmap}).
The profile is shown in Fig.~\ref{fig:i_vertical}.
We characterize the profile by an exponential law: $I(R) = I_0~e^{-|z|/z_0}$, 
where $|z|$ is the vertical distance from the center, $z_0$ is the scale height and $I_0$ is the central intensity.
A constant intensity $I_b$ is included in the fit to account for the discrepancy
between the local background and the subtracted ``blank-sky'' background, and it
turns out to be negligible. 
The results are listed in Table~\ref{tab:inten_fit}.
The best-fit model is also plotted in Fig.~\ref{fig:i_vertical} (black dashed curve).
There is an excess 
over the best-fit model at a vertical distance of $\sim$1$^\prime$-1\farcm5, 
which can be attributed to the extraplanar features apparent in Fig.~\ref{fig:2dmap}.  
The total count rate produced by this extraplanar 
excess is $\sim 5.2 \times10^{-3}~{\rm~cts~s^{-1}}$.

The unresolved emission should consist of two components: 1) 
emission from truly diffuse gas, and 2)
collective discrete contributions from sources below our detection limit
plus some
residual counts spilled outside our source removal regions.
To constrain the source component,
we assume that it follows the distribution of the NIR K-band light of the galaxy,
which can be determined from the 2MASS K-band map (Jarrett et al.~2003).
We remove from the map bright foreground stars 
and convolve it with the PSF of the PN.
Circular regions used for removing the discrete X-ray sources 
are also excluded from the map.
The K-band vertical profile is then produced in the same manner as for the X-ray profile.
To compare the two profiles,
we normalize the K-band profile to match the X-ray intensity at the center (Fig.~\ref{fig:i_vertical}).
This requires a normalization factor of 3.0$\times10^{-3}{\rm~cts~s^{-1}~arcmin^{-2}}/({\rm MJy~sr^{-1}})$.
Within a vertical distance of 1$^\prime$, the K-band profile traces the 
X-ray profile quite well.  An
excess over the K-band profile, however, remains at a vertical distance of 
$\sim$1$^\prime$-1\farcm5. 
This is evidence for the presence of extraplanar X-ray-emitting gas.

\begin{figure}[!htb]
  \centerline{
      \epsfig{figure=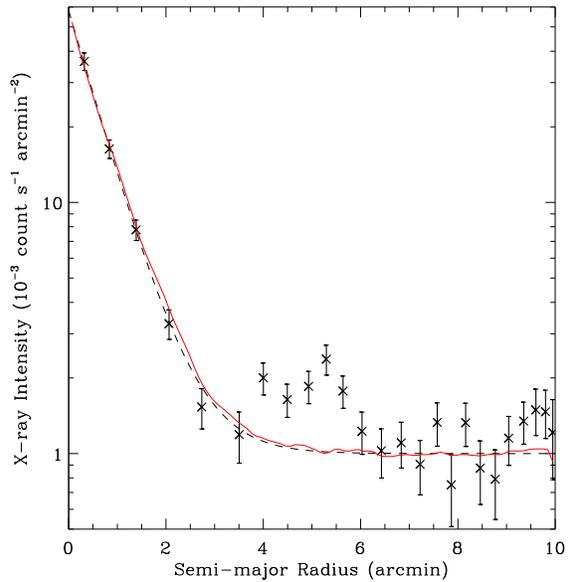,width=0.5\textwidth,angle=0}    
    }
  \caption{Similar to Fig.~\ref{fig:i_vertical} but for the radial profile,
generated with the elliptical photometry (see text).
A constant of $10^{-3}{\rm~cts~s^{-1}~arcmin^{-2}}$ has been added
to all data points to avoid negative values improper for a logarithmic plot.}
 \label{fig:i_radial}
\end{figure}

We also construct a 0.5-2 keV radial intensity profile for the unresolved emission. 
Elliptical photometry is applied, with a minor-to-major axis ratio of 0.29 and a
position angle of 113$^{\circ}$ (Jarrett et al.~2003).
We again fit the radial profile with an exponential law and list the  
results in Table~\ref{tab:inten_fit}. 
The X-ray profile together with the best-fit model is plotted in Fig.~\ref{fig:i_radial}.
Also plotted is the normalized 2MASS K-band radial profile produced in the same way as
is done for the X-ray profile.
The K-band and X-ray profiles closely trace with each other within a semi-major radius of $\sim3^\prime$. 
At semi-major radii $\sim$4$^\prime$-6$^\prime$, corresonding to semi-minor radii 
$\sim$1\farcm1-1\farcm7, 
a bump is present, again most likely due to the extraplanar features. 
The total count rate of this bump is consistent with the extraplanar excess 
seen in the vertical intensity profile.

\subsubsection{Spectral properties}
Guided by the X-ray morphology (Fig.~\ref{fig:2dmap}),
a galactocentric 5$^\prime$ $\times$ 3$^\prime$ ellipse with a position angle of 113$^{\circ}$ 
is adopted to extract the spectra of unresolved emission for individual detectors.
The limited number of counts prevents us from further dividing the spectra according to different
regions of interest, e.g., the disk and the halo.
Discrete sources are removed in the way as described above,
except for the nucleus, for which we exclude a circular region with a 
radius of 40$^{\prime\prime}$ ($\sim$87\% EER) 
to further reduce the contamination from the hard nuclear emission.     
We use the same background spectra as applied in \S~{\ref{subsec:ps}}.

\begin{figure}[!htb]
  \centerline{
      \epsfig{figure=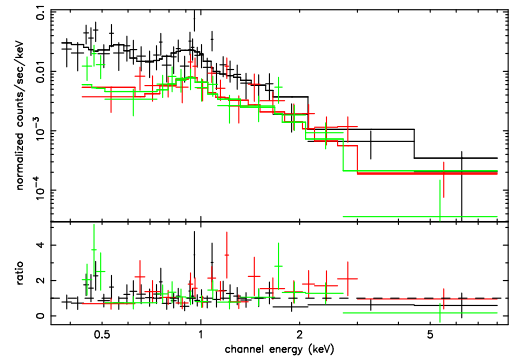,width=0.4\textwidth,angle=0}
    }
  \caption{EPIC spectra of unresolved X-ray emission (Black: PN spectrum; Red: MOS1 spectrum; Green: MOS2 spectrum) of NGC~2613 
and the best-fit {\sl wabs(PL+2APEC)} model.  The spectra are adaptively binned to achieve a background-subtracted 
signal-to-noise ratio better than 2. The lower panel shows the data-to-model ratios.
   }
 \label{fig:diffuse_spec}
\end{figure}

We jointly fit the PN, MOS1 and MOS2 spectra in the 0.3-8 keV range (Fig.~\ref{fig:diffuse_spec}).
Overall, the spectra are much softer than those from the nuclear region,
showing clear features at $\sim$~0.9 keV, corresponding to
the Fe L-shell complex, and at $\sim$~0.5 keV.
These features further indicate the presence of diffuse hot gas.
At energies above 2 keV the spectra are dominated by a collective contribution from 
unresolved discrete sources, most likely LMXBs (see \S~\ref{subsec:cxs}).
We account for this contribution with a power-law 
(PL) with a photon index fixed at 1.56, as found by Irwin, Athey \& Bregman (2003)
for the accumulated spectra of galactic LMXBs.
This PL, combined with a thermal plasma component (APEC) is then used to fit the spectra.
Foreground absorption is again required to be at least the Galactic value. 
The fit is initialized by fixing the plasma metal abundance at solar.
The model, however, yields a poor fit to the spectra, in particular failing
to simultaneously account for the features at $\sim$0.9 keV and $\sim$0.5 keV.
By allowing the metal abundance to be fitted, we obtain a statistically better fit 
but the resulting abundance
is low ($\lesssim$ 0.05 solar). Such an extremely sub-solar abundance is unphysical 
and practically often encountered in the X-ray spectral analysis for  
galaxies (e.g., NGC~253, Strickland et al.~2002; NGC~4631, Wang et al. 2001).
We thus add to the model a second APEC component. A two-temperature plasma
is effective in characterizing the diffuse spectra of some star-forming disk galaxies 
(e.g., Strickland et al.~2004a; T$\ddot{u}$llmann et al.~2006).
The abundances for both thermal components are fixed at solar.
The fit is acceptable, resulting in a cool component with $kT$ $\sim$0.08 keV
and a hot component with $kT$ $\sim$0.8 keV.
The 0.3-10 keV X-ray luminosities are 
5.2, 2.6 and 1.1${\times}10^{39}{\rm~ergs~s^{-1}}$ for the three components, 
namely the discrete sources, the low temperature gas and the high temperature
gas, respectively. 
The above results are summarized in Table~\ref{tab:diffuse_fit}.
We adopt the two-temperature fit as the best-fit model in the following and plot it in 
Fig.~\ref{fig:diffuse_spec}.

\section{Discussion} {\label{sec:discussion}}
\subsection{The nature of the nuclear X-ray emission} {\label{subsec:nucleus}}
Our two-component spectral fit for the nucleus (\S~\ref{subsec:ps}; Table \ref{tab:nucleus_fit}) 
indicates that the intrinsic neutral hydrogen 
column density is $\sim$$1.2{\times}10^{23}{\rm~cm^{-2}}$. 
This is much higher than 
the beam-averaged \HI column density
of $\sim$$2\times10^{21}{\rm~cm^{-2}}$ found by Chaves
\& Irwin (2001), but it is typical of 
 values found for molecular circumnuclear disks.  Ott et al. (2001), for example,
find a molecular column density of order $10^{23}{\rm~cm^{-2}}$ for NGC~4945.
Molecular data are
not yet available for NGC~2613, but our results suggest that a substantial molecular
component should be present in this galaxy.  

For the nuclear component, the modelled intrinsic flux given in 
Table~\ref{tab:nucleus_fit} leads to an
intrinsic X-ray luminosity of $\sim$$3.3{\times}10^{40}{\rm~ergs~s^{-1}}$ 
in the 0.3-10 keV range.  The photon index of this component
is $\sim$2, a typical value
found in the X-ray spectra of AGNs (e.g. Pellegrini, Fabbiano \& Kim 2003).
No radio core was detected by Irwin, Saikia \& English (2000), putting
a 3$\sigma$ 
upper limit of $4.5\times10^{27}{{\rm ~ergs~s^{-1}~Hz^{-1}}}$ on the
radio spectral power at 1.425 GHz within the same 16$^{\prime\prime}$ region.
Using the above luminosity over the 0.3-10 keV range, we derive
 an upper limit of $\alpha\,=\,0.62$ on
the energy spectral index ($S_\nu\,\propto\,\nu^{-\alpha}$) between the radio
and X-ray bands.  Although the X-ray nucleus is heavily obscured, these values
nevertheless suggest that the energy spectral index is likely flat or
possibly rising
at the low frequencies, a fact
again consistent with the interpretation of the nuclear source
as an AGN.  Thus, we conclude that the nuclear X-ray source represents an
AGN in this galaxy, the first evidence that this is the case.

The non-nuclear component, characterized by the second power-law
(PL2), shows a photon index 
of $\sim1.7$
and an intrinsic 0.3-10 keV luminosity of $3.4{\times}10^{39}{\rm~ergs~s^{-1}}$. 
Irwin et al.~(2003) showed that the accumulated spectra of LMXBs in early-type 
galaxies can be uniformly described by a power-law model with a best-fit photon index of $1.56\pm{0.02}$.
By using a sample of nearby galaxies of various morphological types,
Gilfanov (2004) studied the relation between the collective luminosity of LMXBs 
and the K-band luminosity, $L_K$, of the underlying stellar content.
He found that, for LMXBs with luminosity higher than $10^{37}{\rm~ergs~s^{-1}}$,
their collective luminosity $L_X$ = ${\rm(3.3-7.5)}{\times}10^{39}{\rm~ergs~s^{-1}}L_K/10^{11}L_{\odot,K}$. 
Assuming that the spatial distribution of LMXBs follows that of the K-band star 
light (Jarrett et al.~2003),
we estimate that the collective luminosity of LMXBs within the 16$^{\prime\prime}$ circle 
is ${\sim}$${\rm(1.8-4.2)}{\times}10^{39}{\rm~ergs~s^{-1}}$.
Thus the non-nuclear component is consistent with the collective
emission of unresolved LMXBs.
We note that high-mass X-ray binaries (HMXBs) are expected to be present in star-forming 
disk galaxies and their composite
spectral properties are somewhat similar to that of the LMXBs, thus the collective contribution
of HMXBs may also be partly responsible for the non-nuclear component. 
We show below that in NGC~2613 the relative contribution of HMXBs is small as compared to that
of LMXBs.

\subsection{The collective X-ray emission of discrete sources} {\label{subsec:cxs}}
It is known that X-ray binaries, including LMXBs and HMXBs, dominate
the X-ray source populations with luminosities $\gtrsim 10^{35}{\rm~ergs~s^{-1}}$ in galaxies.
Owing to their distinct evolution time-scales, the numbers and thus the 
collective contributions of long-lived LMXBs
and short-lived HMXBs to the X-ray emission of a galaxy are expected to be proportional to 
its stellar mass and star formation rate (SFR), respectively.
Colbert et al.~(2004) analyzed {\sl Chandra} observations of X-ray sources in
a sample of nearby galaxies of various morphological types and SFRs.
They found that the collective X-ray luminosity of point sources $L_{XP}$ is linearly correlated
with the total stellar mass $M_{\star}$ and the SFR of the host galaxy as
\begin{eqnarray}
L_{XP}~(\rm{ergs~s^{-1}})~=~(1.3\pm{0.2})\times 10^{29}~M_{\star}~(\rm{M_{\odot}}) \nonumber \\   
+ (0.7\pm{0.2})\times 10^{39}~\rm{SFR}~(\rm{M_{\odot}~yr^{-1}}). 
\label{eq:LXP}
\end{eqnarray}

We use this relation to assess the relative importance of LMXBs and HMXBs in contributing to 
the X-ray emission of NGC~2613.  The total stellar mass can be estimated from 
the K-band luminosity $L_K$ and the $B-V$ color index via (Bell \& de Jong 2001)
\begin{equation}
{\rm log}(M_{\star}/L_K) = -0.692 + 0.652 (B-V),
\label{eq:Mstar}
\end{equation}
where $L_K$ is in units of the K-band Solar luminosity.
The SFR can be estimated from the far-infrared (FIR) luminosity $L_{FIR}$ via (Kennicutt 1998)
\begin{equation}
\rm{SFR}~=~4.5\times10^{-44}~L_{FIR}~(\rm{ergs~s^{-1}}). 
\label{eq:SFR}
\end{equation}  
$L_{FIR}$ is measured according to (Lonsdale, Helou \& Good 1989)
\begin{equation}
L_{FIR} = 3.1\times10^{39}~D^2~(2.58~S_{60}+S_{100}),
\label{eq:LFIR}
\end{equation}
where $D$ is the distance of the galaxy in units of Mpc, $S_{60}$ and $S_{100}$
are the flux densities in units of Jy at 60~${\mu}$m and 100~${\mu}$m, respectively.
With the available photometric data for NGC~2613 (Table \ref{tab:N2613}), 
we estimate that the total stellar mass is $2.1\times10^{11}{\rm~M_{\odot}}$ and 
the SFR is $4.2{\rm~M_{\odot}~yr^{-1}}$. 
Based on Eq.~(\ref{eq:LXP}), the contributions of LMXBs and HMXBs to the collective X-ray emission of discrete
sources is $\sim2.7\times10^{40}{\rm~ergs~s^{-1}}$ and $\sim2.9\times10^{39}{\rm~ergs~s^{-1}}$, respectively,
with the latter being about 10\% of the former. 

In the disk of NGC~2613, we find that 
the 0.5-2 keV unresolved emission is spatially correlated with the K-band
star light.  Therefore, the
normalization factor for the K-band profile (\S~\ref{subsubsec:spat_anal}) 
should represent the collective X-ray emissivity of the underlying old stellar population.
Using the power-law model given by Irwin et al.~(2003)
for the accumulated spectrum of LMXBs,
we convert the observed 0.5-2 keV count rate into the intrinsic luminosity in the
0.3-10 keV band).
The K-band flux density is also converted into intrinsic luminosity according to the 2MASS K-band photometry.
The normalization factor, 3.0$\times10^{-4}{\rm~cts~s^{-1}~arcmin^{-2}}/({\rm MJy~sr^{-1}})$,
is then equivalent to an X-ray emissivity of
$L_X = 4.2{\times}10^{39}{\rm~ergs~s^{-1}}L_K/(10^{11}L_{\odot,K})$,
or a luminosity ratio of $L_X/L_K \sim 7.5\times10^{-4}$.
Gilfanova (2004) found that
the collective X-ray luminosity of galactic LMXBs is related to the underlying K-band
luminosity following $L_X = (3.3-7.5) {\times}10^{39}{\rm~ergs~s^{-1}}L_K/(10^{11}L_{\odot,K})$,
i.e., a luminosity ratio of $L_X/L_K \sim (5.8-13.2)\times10^{-4}$.
Therefore the collective X-ray emissivity of unresolved discrete sources inferred for NGC~2613
is consistent with that of the galactic LMXBs, 
and we conclude that the collective X-ray emission of LMXBs dominates the soft emission of NGC~2613
in its disk region.  

\subsection{The origin of extraplanar gas}
Presence of diffuse gas in NGC~2613 is evident by the soft X-ray excess over
the K-band light.
We consider two possible origins of the diffuse gas:
1) the continuously accreted IGM (Toft et al. 2002) and
2) the outflow from the galactic disk (Irwin \& Chaves 2003). 

\subsubsection{An accreted gaseous halo?}
Toft et al.~(2002) calculated global X-ray properties (e.g., luminosity, effective temperature
and intensity distribution) of hot gaseous halos, based on
their simulated galaxies. The predicted luminosity strongly depends on
the circular speed of the host galaxy. The most massive galaxies in their simulations
have circular speeds similar to that of NGC~2613 ($\sim$ $300{\rm~km~s^{-1}}$).
The predicted 0.2-2 keV luminosity for such a galaxy is
$\sim8\times10^{40}{\rm~ergs~s^{-1}}$ (Fig.~3 in Toft et al.~2002).
From our best-fit spectral models of the spectra of unresolved emission, 
we derive an intrinsic 0.2-2 keV luminosity of $\sim$8$\times10^{39}{\rm~ergs~s^{-1}}$ 
for the sum of the thermal and power-law
components, and $\sim$6$\times10^{39}{\rm~ergs~s^{-1}}$ for the thermal components only.
We note that the unresolved emission outside our spectral extraction region contributes little to the
total luminosity.
The simulated luminosity of gas emission by Toft et al.~(2002) is
at least an order of magnitude higher than the observed value for NGC~2613.
We therefore conclude that the simulations as presented by Toft et al.~(2002)
substantially over-predict the X-ray emission from the cooling inflow of the IGM,
if this is what is occurring in NGC~2613.

This over-prediction is related to the so-called over-cooling problem in current theories of
galaxy formation.  
We speculate that the
over-cooling problem is a result of an inappropriate treatment
of stellar and/or AGN feedback. For example, the mechanical energy input from
Type Ia supernovae (SNe) is typically not included in galaxy formation simulations, partly
due to the difficulty in treating the astrophysics related to gaseous flows. 
Qualitatively,
Type Ia SNe, which tend to occur in low-density hot environments, provide an especially 
effective mechanism for large-scale distributed heating, required to reduce
the cooling of gas in galactic bulges and halos (Tang \& Wang 2005). 
Massive stars in galactic disks may also serve as sources of mechanical energy 
that could produce outflows into halos and help slow down the cooling of the accrected gas. 
For example,
with a star formation rate of $\sim4.2~{\rm M_{\odot}~yr^{-1}}$
for stars between 0.1 and 100 ${\rm M_{\odot}}$,
and assuming a Salpeter IMF and that stars with mass $> 8 {\rm~M_{\odot}}$ become
core-collapse SNe,
the rate of total energy release from the star-forming regions of NGC~2613
is $L_{SNII}\sim1.0\times10^{42} {\rm~ergs~s^{-1}}$.
Our spatial and spectral analyses suggest that the extraplanar gas
is responsible for the thermal emission (\S~\ref{subsec:diffuse}) and has a total 0.3-10 keV luminosity
of $\lesssim5\times10^{39}{\rm~ergs~s^{-1}}$.
Thus, SNe can provide enough energy to
explain the X-ray emission of the extraplanar gas in NGC~2613.

\begin{figure*}[!htb]
 \vskip -1cm
 \centerline{
       \epsfig{figure=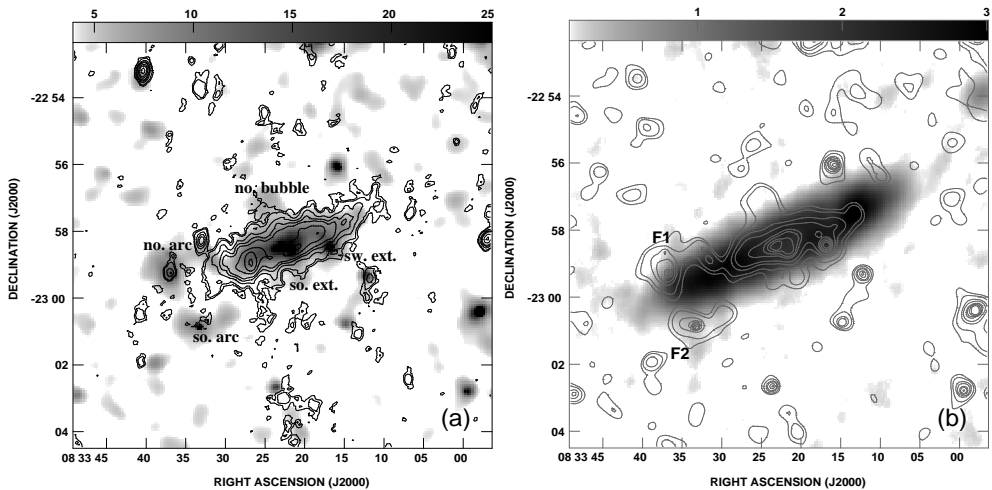,width=0.8\textwidth,angle=0}
    }
    \vskip -0.5cm
 \caption{(a) 
VLA C+D configuration continuum contours overlaid on the same X-ray intensity image (grey scale)
as contoured in Fig.~\ref{fig:2dmap} and in (b) of this figure.
The contour levels are 0.18, 0.27, 0.56, 0.84, 1.1, 1.7, 2.3, 3.2~mJy~beam$^{-1}$ 
and the beam is $22^{\prime\prime}{\times}15^{\prime\prime}$ at a position angle of $-8\fdg2$.
A few X-ray extraplanar features are labelled (see text).
(b) The same X-ray intensity contours as in Fig.~\ref{fig:2dmap}
overlaid on a greyscale image of the total intensity VLA C+D configuration
\HI map.
The grey scale range (shown with a square root transfer function) 
 is in units of $10^3$ Jy beam$^{-1}$ m s$^{-1}$ and the beam is
$47^{\prime\prime}{\times}32^{\prime\prime}$ at a position angle of $-8\fdg2$. F1 and F2 refer
to two \HI extensions identified by Chaves \& Irwin (2001).
}
 \label{fig:overlay_map}
\end{figure*}

\subsubsection{Multiwavelength extraplanar features} {\label{subsubsec:correlations}}
In Fig.~\ref{fig:overlay_map}a and b, we compare the X-ray emission with
the radio continuum emission and \HI total intensity emission, respectively.
Of the two radio images,
the radio continuum morphology more closely resembles the X-ray morphology in
the sense that: a) the north bubble  has a
radio continuum counterpart; b) the south extension
also has a radio continuum counterpart; c) the south-west feature
shows a small radio continuum protrusion; and d)
the peaks of the large eastern extensions (north and south) also show radio emission.

The \HI total intensity map shown here does not show all of the extended features 
identified by Chaves \& Irwin
(2001), but two of their features, F1 and F2 clearly extend above and below the galactic plane
and are labelled in Fig.~\ref{fig:overlay_map}b.  These two features might be related with
the northern and southern arc of the eastern extensions seen in the X-ray.

It is not wise to read too much into these correlations, given the limited S/N of the maps.
However, the relationship with the radio continuum is sufficiently strong that the
X-ray emission in the extraplanar features, representing hot diffuse gas, 
is very likely associated with the
radio continuum emission which represents predominantly the non-thermal component.

\begin{figure}[!htb]
  \vskip -1.5cm
   \centerline{
      \epsfig{figure=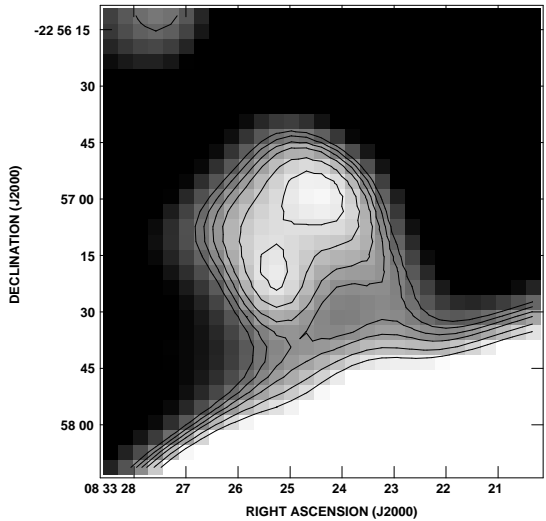,width=0.5\textwidth,angle=0}
    }
  \vskip -2cm
  \caption{Blow-up of the X-ray-emitting bubble to the north of the nucleus. 
The same X-ray intensity image as in Fig.~\ref{fig:2dmap} is used.
Contour levels are at 5, 5.2, 5.4, 5.6, 5.9, and 6.2
$\times$10$^{-3}{\rm~cts~s^{-1}~arcmin^{-2}}$.
}
 \label{fig:bubble_map}
\end{figure}

We further consider 
the energetics of a specific feature, namely 
the `north bubble', which is the only extraplanar feature
that can be cleanly isolated from the ambient emission. 
Guided by Fig.~\ref{fig:bubble_map}, we approximate the volume occupation of the bubble
by a cylinder with 1$^\prime$ in diameter and 0\farcm8 in height, the center
of which is 1\farcm2 above the galactic center.
Hence the volume of the bubble is $\sim2.7\times10^2{\rm~kpc^3}$. 
We find a total 0.5-2 keV count rate of 2.4$\times10^{-3}{\rm~cts~s^{-1}}$ within the bubble. 
In the best-fit model to the spectra of unresolved emission, 
the high and low temperature components predict
a 0.5-2 keV count rate of 7.5$\times10^{-3}{\rm~cts~s^{-1}}$ and
2.0$\times10^{-3}{\rm~cts~s^{-1}}$, respectively. 
Therefore the north bubble
is unlikely to be due to the low temperature component alone.
Instead, it could be dominated by the high temperature component.
Taking an effective temperature of $\sim$ 0.8 keV,
we estimate the mean density of the bubble to be $\sim\eta^{-1/2}\times10^{-3}{\rm~cm^{-3}}$,
where $\eta$ is the filling factor of the hot gas inside the bubble.
The total thermal energy of the bubble is $E_{th}\simeq 3.6\eta^{-1/2}\times10^{55}{\rm~ergs~s^{-1}}$,
and the work done to steadily lift up the bubble against gravity is
$E_g \simeq 1.7\eta^{-1/2}\times10^{55}{\rm~ergs~s^{-1}}$,
given the gravitational potential introduced by
the exponential disk of the galaxy (Irwin \& Chaves 2003). 
Given the morphology and the position of the bubble,
we speculate that it was produced near the nuclear region, either by a nuclear starburst or the AGN.
If the bubble's total amount of thermal and gravitational energy is obtained from a starburst,
it takes a time of 
$\tau\simeq(E_{th}+E_g)/(fL_{SNII})\simeq1.7\eta^{-1/2}\times10^7{\rm~yr}$
to form the present structure, where $f$ is a geometrical factor taken to be 0.1
to reflect a fractional star formation rate of the central 1 kpc in the disk.  
This timescale is typical for massive stars to become SN explosions.
On the other hand, assuming the flux density of the AGN follows $S_\nu\,\propto\,\nu^{-\alpha}$ between the radio
and X-ray bands, the total bolometric luminosity over this frequency range is 
$\sim 4.5\times10^{40} {\rm~ergs~s^{-1}}$ with $\alpha=0.62$ adopted (\S~\ref{subsec:nucleus}).
It is uncertain what fraction of the AGN energy can be taken to energize
the ambient gas, but we consider the AGN might also be capable of producing this feature. 
For example, the locations
of the north bubble and south extension immediately on either side of the nucleus
are reminiscent of 
extraplanar loops or lobes seen in nuclear outflow galaxies like 
NGC~3079 (e.g. Cecil et al. 2002) which is known to have an AGN.


\section{Summary} {\label{sec:summary}}
We have analyzed an \xmm~observation of the massive edge-on Sb galaxy NGC~2613.
We find a deeply embedded AGN in this galaxy. The X-ray spectrum of this AGN 
can be characterized by a power-law model with a photon-index of $\sim$2
and a 0.3-10 keV intrinsic luminosity
of 3.3$\times10^{40}{\rm~ergs~s^{-1}}$. Linking the X-ray spectral properties of the AGN
with the current upper limit at radio frequencies indicates a spectral flattening
of the AGN at low frequencies.

The 0.5-2 keV unresolved X-ray emission is found to closely trace the near-IR emission in the disk
region, and the X-ray to near-IR luminosity ratio is consistent with
that inferred from galactic LMXBs. 
These two facts together
indicate that the bulk of the unresolved emission is produced by the old stellar population
of the galaxy, predominantly LMXBs. 

A few extraplanar diffuse X-ray features are present in addition to 
the collective emission from discrete sources traced by the near-IR light.
These features can be explained by the presence of hot gas,
which can be spectrally characterized by a two-temperature plasma with
$kT$ $\sim$0.08 keV and $\sim$0.8 keV.
The total X-ray luminosity of hot gas is at least
an order of magnitude lower than that predicted by current simulations
of IGM accretion based on disk galaxy formation models. 
Thus the extraplanar features are very
unlikely to result from IGM accretion.

Instead, morphologically most of these extraplanar features have extended radio counterparts,
which are believed to arise from disk-related events.
Also, energetically the extraplanar features can be generated by either supernova explosions
or the AGN, the latter possibly related to the bubbles above and below the
nucleus.
Therefore, we conclude that the extraplanar features are most likely formed
from outflows from the galactic disk.

Our observation  
suggests that a proper inclusion of galactic feedback is essential, not only
 to understanding
galaxy formation, but also to its continued evolution.  NGC~2613 and galaxies
like it provide nearby laboratories that may help to understand the over-cooling
problem existing in current galaxy formation simulations.

For J. I., this work has been supported by the Natural Sciences and
Engineering Research Council of Canada.

\begin{deluxetable}{lr}
\tablecaption{Basic information of NGC~2613}
\tablewidth{0pt}
\tablehead{
\colhead{Parameter} &
\colhead{NGC~2613}}
\startdata
Morphology$^a$     \dotfill & SA(s)b \\
Optical size$^a$\dotfill & $7\farcm2\times1\farcm8$ \\
Inclination angle$^b$  \dotfill & 79$^\circ$\\
Position angle$^c$ \dotfill & 113$^{\circ}$ \\
Center position$^a$ \dotfill & R.A.~$08^{\rm h} 33^{\rm m} 22\fs84$  \\
~~ (J2000) \dotfill & Dec.~$-22^\circ 58^\prime 25\farcs2$ \\
B-band magnitude $^a$  \dotfill &  11.16 \\
V-band magnitude $^a$ \dotfill & 10.25 \\
K-band magnitude $^c$ \dotfill & 6.82 \\
60 ${\mu}$m flux (Jy)$^d$ \dotfill & 7.48 \\
100  ${\mu}$m flux (Jy)$^d$ \dotfill & 25.86\\
Circular speed (${\rm km~s^{-1}}$)$^b$\dotfill  & $304$\\
Distance (Mpc)$^b$ \dotfill & $25.9$ \\
\dotfill & ($1^\prime~\cor~$~7.53kpc) \\
Redshift$^a$ \dotfill & $0.00559$\\
Galactic foreground $N_{\rm \HI}$ ($10^{20}{\rm~cm^{-2}}$)$^e$\dotfill &$6.8$ \\
\enddata
\tablerefs{
 $a.~$NED;
 $b.~$Chaves \& Irwin (2001);
 $c.~$Jarrett et al. (2003);
 $d.~$Sanders et al. (2003);
 $e.~$Dickey \& Lockman (1990).
}
\label{tab:N2613}
\end{deluxetable}
\vfill
\eject

\begin{deluxetable}{lrrrrrrrr}
  \tabletypesize{\footnotesize}
  \tablecaption{{\sl XMM-Newton} Source List \label{tab:pn_source_list}}
  \tablewidth{0pt}
  \tablehead{
  \colhead{Source} &
  \colhead{XMMU Name} &
  \colhead{$\delta_x$ ($''$)} &
  \colhead{CR $({\rm~cts~ks}^{-1})$} &
  \colhead{HR} &
  \colhead{HR1} &
  \colhead{Flag} \\
  \noalign{\smallskip}
  \colhead{(1)} &
  \colhead{(2)} &
  \colhead{(3)} &
  \colhead{(4)} &
  \colhead{(5)} &
  \colhead{(6)} &
  \colhead{(7)} 
  }
  \startdata
   1 &  J083228.57-225840.5 &  5.2 &$     5.58  \pm   1.45$& --& --& B, S \\
   2 &  J083231.32-230254.1 &  4.5 &$     6.39  \pm   1.57$& --& --& S, B \\
   3 &  J083233.69-225922.0 &  3.2 &$    70.77  \pm   4.21$& $-0.31\pm0.09$ & $ 0.03\pm0.07$ & B, S, H \\
   4 &  J083235.26-225805.9 &  3.3 &$    27.84  \pm   2.54$& $-0.18\pm0.15$ & $ 0.02\pm0.10$ & B, S, H \\
   5 &  J083236.47-225734.6 &  4.0 &$     6.41  \pm   1.37$& --& $-0.34\pm0.20$ & S, B \\
   6 &  J083238.91-225942.2 &  3.7 &$     6.53  \pm   1.35$& --& $-0.89\pm0.15$ & B, S \\
   7 &  J083241.40-225809.7 &  3.4 &$     6.48  \pm   1.24$& --& --& B, S \\
   8 &  J083241.81-225523.7 &  4.0 &$     5.48  \pm   1.26$& --& --& B, S \\
   9 &  J083241.95-230446.3 &  3.6 &$    10.06  \pm   1.69$& --& $ 0.08\pm0.20$ & B, S \\
  10 &  J083246.75-230209.9 &  3.6 &$     4.40  \pm   1.10$& --& --& B, S \\
  11 &  J083247.39-230258.2 &  2.1 &$    55.57  \pm   3.23$& $-0.10\pm0.08$ & $ 0.13\pm0.07$ & B, S, H \\
  12 &  J083248.13-225441.8 &  3.8 &$     4.81  \pm   1.08$& --& --& B, S \\
  13 &  J083249.89-230206.3 &  3.1 &$     6.50  \pm   1.20$& --& $-0.57\pm0.15$ & S, B \\
  14 &  J083252.35-225024.1 &  4.8 &$     5.76  \pm   1.33$& --& --& B \\
  15 &  J083252.35-230038.8 &  3.4 &$     3.55  \pm   0.90$& --& --& B \\
  16 &  J083254.56-225236.7 &  3.5 &$     7.51  \pm   1.31$& --& $-0.06\pm0.20$ & B, S \\
  17 &  J083255.26-224605.1 &  6.3 &$    13.38  \pm   2.65$& --& --& B, S, H \\
  18 &  J083255.42-230253.4 &  2.3 &$     8.39  \pm   1.25$& $-0.02\pm0.20$ & $ 0.39\pm0.18$ & B, S, H \\
  19 &  J083257.98-230023.9 &  1.5 &$    18.91  \pm   1.81$& $-0.11\pm0.14$ & $ 0.11\pm0.11$ & B, S, H \\
  20 &  J083258.24-225202.7 &  2.8 &$    16.52  \pm   1.76$& $-0.24\pm0.19$ & $-0.10\pm0.12$ & B, S, H \\
  21 &  J083259.34-230245.2 &  2.7 &$     6.17  \pm   1.28$& --& --& B, S \\
  22 &  J083305.75-224832.5 &  3.5 &$   653.78  \pm  11.91$& $ 0.04\pm0.03$ & $ 0.10\pm0.02$ & B, S, H \\
  23 &  J083309.78-224757.0 &  5.2 &$     6.76  \pm   1.92$& --& --& S \\
  24 &  J083310.11-225621.4 &  2.7 &$     3.25  \pm   0.75$& --& --& B, H \\
  25 &  J083312.19-225918.9 &  2.1 &$     4.13  \pm   0.77$& --& $-0.42\pm0.19$ & B, S \\
  26 &  J083312.86-225203.9 &  2.6 &$     8.90  \pm   1.26$& --& $-0.58\pm0.12$ & B, S \\
  27 &  J083313.50-224816.7 &  4.6 &$     6.03  \pm   1.49$& --& --& B, H \\
  28 &  J083313.96-225013.6 &  4.0 &$     4.57  \pm   1.11$& --& --& B \\
  29 &  J083314.71-230043.2 &  2.5 &$     2.97  \pm   0.71$& --& --& S, B \\
  30 &  J083315.02-225752.9 &  3.1 &$     1.78  \pm   0.60$& --& --& S \\
  31 &  J083315.94-225604.2 &  2.0 &$     6.95  \pm   1.07$& --& $-0.07\pm0.16$ & B, S \\
  32 &  J083316.82-225827.7 &  1.2 &$    11.60  \pm   1.20$& $-0.15\pm0.14$ & $ 0.33\pm0.12$ & B, S, H \\
  33 &  J083318.20-225211.3 &  2.4 &$     8.95  \pm   1.26$& --& $-0.62\pm0.11$ & S, B \\
  34 &  J083322.83-225825.9 &  0.9 &$    19.96  \pm   1.60$& $ 0.48\pm0.09$ & $-0.03\pm0.13$ & B, H, S \\
  35 &  J083324.91-224505.1 &  5.5 &$    20.16  \pm   2.69$& --& $ 0.14\pm0.14$ & B, S \\
  36 &  J083325.49-225843.6 &  2.2 &$     3.98  \pm   0.81$& --& --& B, S \\
  37 &  J083327.32-230103.1 &  2.6 &$     2.53  \pm   0.69$& --& --& B, H \\
  38 &  J083327.49-225851.7 &  2.4 &$     2.95  \pm   0.72$& --& --& B \\
  39 &  J083329.80-230715.8 &  3.6 &$     4.28  \pm   1.13$& --& --& S, B \\
  40 &  J083331.06-225006.6 &  3.5 &$     7.46  \pm   1.37$& --& $-0.23\pm0.20$ & B, S \\
  41 &  J083333.15-230050.8 &  1.9 &$     7.04  \pm   1.17$& --& $-0.07\pm0.20$ & B, S \\
  42 &  J083334.38-230542.1 &  2.1 &$     9.12  \pm   1.40$& --& $-0.43\pm0.16$ & B, S \\
  43 &  J083336.73-225922.2 &  2.9 &$     2.76  \pm   0.85$& --& --& S \\
  44 &  J083337.40-225124.4 &  3.8 &$     3.27  \pm   0.99$& --& --& S \\
  45 &  J083338.72-230155.1 &  2.5 &$     3.82  \pm   0.88$& --& --& B, S \\
  46 &  J083340.39-230529.7 &  3.5 &$     4.32  \pm   1.07$& --& --& B, S \\
  47 &  J083340.73-225056.0 &  4.0 &$     4.72  \pm   1.18$& --& --& B \\
  48 &  J083342.87-225030.0 &  5.0 &$     3.73  \pm   1.14$& --& --& S \\
  49 &  J083350.48-231150.8 &  5.2 &$     9.28  \pm   2.19$& --& --& B, S \\
  50 &  J083351.30-225656.7 &  4.3 &$     3.31  \pm   1.04$& --& --& S \\
  51 &  J083352.03-225306.0 &  3.0 &$    10.61  \pm   1.63$& --& $-0.33\pm0.15$ & B, S \\
  52 &  J083352.44-230516.5 &  3.3 &$     6.49  \pm   1.38$& --& --& B, S \\
  53 &  J083352.84-231135.6 &  5.4 &$     7.58  \pm   2.05$& --& --& B, S \\
  54 &  J083353.06-225728.9 &  2.5 &$     7.78  \pm   1.32$& --& $-0.16\pm0.18$ & B, S \\
  55 &  J083354.10-230211.9 &  3.1 &$     5.35  \pm   1.21$& --& --& B, S \\
  56 &  J083356.35-230249.5 &  3.4 &$     5.16  \pm   1.30$& --& --& S, B \\
  57 &  J083358.74-230730.8 &  4.3 &$     7.31  \pm   1.71$& --& --& B, S \\
  58 &  J083359.63-230915.8 &  4.0 &$    25.05  \pm   3.05$& --& $-0.06\pm0.12$ & B, S \\
  59 &  J083406.25-224808.9 &  5.7 &$    46.66  \pm   4.55$& $-0.05\pm0.14$ & $ 0.38\pm0.11$ & B, S, H \\
  60 &  J083410.07-225808.2 &  4.4 &$     6.46  \pm   1.55$& --& --& B, S \\
  61 &  J083416.92-225508.4 &  5.6 &$     6.17  \pm   1.85$& --& --& S \\
  62 &  J083417.03-230448.5 &  5.8 &$     7.14  \pm   2.04$& --& --& B \\
  63 &  J083422.96-225632.9 &  4.8 &$    67.11  \pm   5.02$& $-0.11\pm0.11$ & $ 0.10\pm0.09$ & B, S, H \\
  64 &  J083426.43-225921.1 &  5.8 &$    11.11  \pm   2.42$& --& --& S, B \\
  65 &  J083428.17-225454.8 &  6.0 &$    26.10  \pm   3.35$& --& $-0.07\pm0.14$ & B, S \\
  66 &  J083428.60-225508.9 &  7.0 &$    18.48  \pm   2.93$& --& $-0.07\pm0.19$ & H \\
  67 &  J083430.98-225843.4 &  6.3 &$    11.55  \pm   2.42$& --& --& B, S \\
\enddata
\tablecomments{The definition of the bands:
0.5--1 (S1), 1--2 (S2), 2--4.5 (H1), and 4.5--7.5~keV (H2). 
In addition, S=S1+S2, H=H1+H2, and B=S+H.
 Column (1): Generic source number. (2): 
{\sl XMM-Newton} X-ray Observatory (unregistered) source name, following the
{\sl XMM-Newton} naming convention and the IAU Recommendation for Nomenclature
(http://cdsweb.u-strasbg.fr/iau-spec.html). (3): Position 
uncertainty (1$\sigma$) calculated from the maximum likelihood centroiding.  (4): On-axis source broad-band count rate --- the sum of the 
exposure-corrected count rates in the four
bands. (5-6): The hardness ratios defined as 
${\rm HR}=({\rm H-S2})/({\rm H+S2})$, and ${\rm HR1}=({\rm S2-S1})/{\rm S}$, 
listed only for values with uncertainties less than 0.2.
(7): The label ``B'', ``S'', or ``H'' mark the band in 
which a source is detected with the most accurate position that is adopted in
Column (3). 
}
  \end{deluxetable}
\vfill

\begin{deluxetable} {lr}
\tablecaption{Spectral fit to the nuclear emission$^a$}
\tablewidth{0pt}
\tablehead{
\colhead{Parameter} &
\colhead{Value}
}
\startdata
$\chi^2$/d.o.f.\dotfill & 64.6/65 \\
$^b$N$_{\HI}$ \dotfill & 6.8 ($<$11.4) \\
$^c$N$_{\HI}$ \dotfill & 12.3$^{+12.3}_{-5.7}$ \\
Photon index (PL1)\dotfill & 2.1$^{+1.8}_{-0.3}$ \\
Photon index (PL2)\dotfill & 1.7$^{+0.5}_{-0.3}$ \\
$^d$f$_{0.3-10}$ (PL1) \dotfill & 41 \\
$^d$f$_{0.3-10}$ (PL2) \dotfill & 4.2
\enddata
\tablecomments{
a.~See text for model description;
b.~Foreground column density in units of $10^{20}{\rm~cm^{-2}}$, minimum sets at the Galactic foreground value of 6.8;
c.~Intrinsic column density in units of $10^{22}{\rm~cm^{-2}}$; 
d. Intrinsic 0.3-10 keV flux in units of 10$^{-14}{\rm~ergs~cm^{-2}~s^{-1}}$.
}
\label{tab:nucleus_fit}
\end{deluxetable}

\begin{deluxetable}{lcc}
\tablecaption{Fit to 0.5-2 keV surface intensity distributions}
\tablewidth{0pt}
\tablehead{
\colhead{Parameter} & 
\colhead{Vertical distribution} &
\colhead{Radial distribution}
}
\startdata
$\chi^2$/d.o.f.\dotfill & 42.4/24 & 65.2/23 \\
$^a$I$_0$ (10$^{-3}$ cts~s$^{-1}$~arcmin$^{-2}$)\dotfill & 15.3$^{+1.8}_{-1.8}$ & 57.0$^{+8.4}_{-9.1}$\\
$^b$$z_0$ (r$_0$) (arcmin)\dotfill & 0.32$^{+0.04}_{-0.04}$  & 0.65$^{+0.06}_{-0.06}$ \\
$^c$I$_b$ (10$^{-3}$ cts~s$^{-1}$~arcmin$^{-2}$)\dotfill & 0.0$^{+1.0}_{-1.0}$  & $^d$0.0 \\ 
\enddata
\tablecomments{
 a.~Central intensity;
 b.~Scale height (length) of the exponential law;
 c.~Local background intensity above the already subtracted ``blank-sky'' background;
 d.~Fixed at the value obtained from the vertical fit.
}
\label{tab:inten_fit}
\end{deluxetable}

\begin{deluxetable}{lcccccc}
\tablecaption{Spectral fit to the unresolved emission}
\tablewidth{0pt}
\tablehead{
\colhead{Model} &
\colhead{N$_{\HI}$$^a$} & 
\colhead{$\alpha$$^b$} &
\colhead{Temperature$^c$} &
\colhead{Abundance} &
\colhead{Flux$^d$} &
\colhead{$\chi^2$/d.o.f.}
}
\startdata
PL+APEC & 6.8 ($<$7.9) & 1.56 & 0.78$^{+0.18}_{-0.14}$ & 1.0$^e$ & 7.5 (PL), 1.3 & 81.1/66\\
PL+APEC & 6.8 ($<$17.9) & 1.56 & 0.84$^{+0.25}_{-0.49}$ & 0.02 ($<$0.05) & 3.2 (PL), 4.4 &  57.9/65 \\
PL+2APEC & 6.8 ($<$21.9) & 1.56 & 0.08$^{+0.12}_{-0.03}$, 0.81$^{+0.19}_{-0.11}$ & 1.0$^e$ & 6.5 (PL), 3.2, 1.4 & 62.5/64 
\enddata
\tablecomments{
 a.~Column density in units of $10^{20}{\rm~cm^{-2}}$, minimum sets at the Galactic foreground value of 6.8; 
 b.~Power-law photon index, fixed at the uniform value for LMXBs found by Irwin et al. (2003).
 c.~In units of keV;
 d.~Intrinsic 0.3-10 keV fluxes in units of 10$^{-14}{\rm~ergs~cm^{-2}~s^{-1}}$;
 e.~In units of solar, fixed.
}
\label{tab:diffuse_fit}
\end{deluxetable}

\end{document}